# Universal stability of two-dimensional traditional semiconductors

Michael C. Lucking, Weiyu Xie[*], Duk-Hyun Choe, Damien West, Toh-Ming Lu, and S. B. Zhang

*Department of Physics, Applied Physics & Astronomy, Rensselaer Polytechnic Institute, Troy, NY 12180, USA*

**Abstract**

Interest in two dimensional materials has exploded in recent years. Not only are they studied due to their novel electronic properties, such as the emergent Dirac Fermion in graphene, but also as a new paradigm in which stacking layers of distinct two dimensional materials may enable different functionality or devices. Here, through first-principles theory, we reveal a large new class of two dimensional materials which are derived from traditional III-V, II-VI, and I-VII semiconductors. It is found that in the ultra-thin limit all of the traditional binary semi-conductors studied (a series of 26 semiconductors) stabilize in a two dimensional double layer honeycomb (DLHC) structure, as opposed to the wurtzite or zinc-blende structures associated with three dimensional bulk. Not only does this greatly increase the landscape of two-dimensional materials, but it is shown that in the double layer honeycomb form, even ordinary semiconductors, such as GaAs, can exhibit exotic topological properties.

[*]co-first author

Until now, most known two-dimensional (2D) materials are derivatives of layered three-dimensional (3D) materials. From a coordination chemistry viewpoint, however, the crystal structure of any 2D system, or thin film, need not be that of bulk. For example, atomic-layer-thin semiconductors exist in the single-layer honeycomb (SLHC) structure such as graphene, silicene, and germanene [1–5] with variable stability. Are these merely happenstances, or do they suggest a universal trend that *all* bulk materials could be synthesized in some form of layered structure? Recent experiment suggests that this may indeed be the case [6], where by using a bilayer graphene as a capping layer, one can grow GaN, a traditional wide-gap 3D semiconductor, into a bilayer on a SiC substrate. This opens the door for engineering layered structures out of conventional binary semiconductors. Additionally, there have been some theoretical indications, in which first-principles calculations have shown that, at least a handful of the binary semiconductors such as GaN and ZnO can be stabilized in the SLHC form, as judged by the lack of imaginary phonon frequencies [1,2,7]. However, unlike graphene, but similar to silicene and germanene, these artificial 2D semiconductors usually buckle due to the chemical difference between A and B elements.

Study of 2D materials has been intense, fueled by the realization of novel properties and quantum physics at confined dimensions. Graphene, for example, exhibits an unusual relativistic Dirac fermion behavior at the Fermi level, giving rise to exceptionally large carrier mobility. Silicene and germanene, while maintaining certain advantages of graphene, offer enhanced spin-orbit coupling (SOC). Hexagonal boron nitride (*h*-BN) is, on the other hand, a 2D insulator, which can be used to support and separate other 2D materials [8]. While SLHCs, other than *h*-BN, are yet to be synthesized, first-principles calculations suggested that they are semiconductors with a band gap typically larger than bulk, which is in startling contrast to other emerging 2D semiconductors, e.g., transition metal dichalcogenides which exhibit intervalley coupling [9–12], and to 2D metals, e.g., unit-cell-layer-thick metallic FeSe films on strontium titanate which exhibit high-temperature superconductivity at a $T_C = 109\ K$ [13].

In this paper, we show by first-principles calculations that many traditional 3D semiconductors can also exist in stable layered forms with structures which are distinct from their three dimensional counterparts. By surveying binary semiconductors, we find that in the ultra-thin limit, their most stable form is neither that of truncated bulk or the SLHC structure, but instead is a double-layer honeycomb (DLHC) structure where individual SLHCs are bound together by dative bonds. Although a dative bond is weaker than a covalent bond, the doubling of the bond density and an elimination of chemically-reactive cation dangling bonds make the DLHC more stable than a truncated bulk. Additionally, multiple-layer DLHCs can also form with pure van der Waals (vdW) interaction between layers. These kinetically and energetically stable DLHCs are rich in novel properties including band inversion in InSb, InAs, GaSb, GaAs, and HgTe and an associated normal (NI) to topological (TI) insulator transition which depends on the number of stacked DLHCs. In the absence of a gap closure, on the other hand, it becomes an alternation in dipole-allowed/forbidden transitions. As exciton binding energy increases significantly in 2D systems, a vanishing band gap can also lead to the formation of exciton insulator, e.g., in HgTe, which has been difficult in 3D systems.

Our calculations are performed using the density functional theory (DFT) within the Perdew-Burke-Ernzerhof (PBE) [14] approximation for exchange-correlation functional and the projector augmented wave (PAW) method [15], as implemented in the VASP code [16]. The vdW interactions are included by using the DFT D3 method [17]. A 600-eV cutoff energy is used for plane wave expansion. The convergence criterion for electronic relaxation is $10^{-6}$ eV. A Γ-centered 11x11 k-point grid in the Brillouin zone (BZ) is used for DLHCs, whereas a Γ-centered 12x12 k-point grid is used for films with bulk structures. For bulk zincblende (ZB) and WZ materials, 12 k-points in the $z$-direction are used. The lattice constants and atomic positions are both optimized until the forces are less than 0.01 eV/Å. For WZ-MnS, the DFT+U method with U = 7 eV is used, which reproduces the experimental band gap (~2.3 eV). In a number of cases, Heyd-Scuseria-Ernzerhof (HSE) calculations with SOC are performed at a k-point grid 5×5×1. To include explicitly excitons in optical transitions, we perform time-dependent Hartree-Fock (TDHF) calculations [18] based on the HSE + SOC+D3 (HSD) results. Z2Pack code [19] is used to calculate the Z2 invariant.

Previous studies showed that a number of binary semiconductors can exist as SLHC, often when either A or B is a first-row element (e.g., B, N, or O) [1,2]. A large number of conventional semiconductors are, however, unstable in the SLHC structure, as evidenced by their imaginary phonon frequencies [1,2,7]. However, here we find that two unstable SLHCs can bind spontaneously to form a stable DLHC, which can also be viewed as a transformation of a 2-monolayer thick truncated bulk (TB) by displacing the topmost-cation layer with respect to the remainder of the slab (see, e.g., Fig. 1a for GaAs) [5]. In a way, the transformation "hides" surface cations by doubling the number of interlayer bonds, as indicated by the dashed lines in Fig. 1a. By symmetry, all interlayer bonds in DLHC are identical (see Fig. 1b). However, they are qualitatively different from original AB bonds, as revealed by the electron localization function (ELF) in Figs. 1c and 1d. Although being more localized to the anion, ELF for TB shows electron localization in the interlayer region. In contrast, for the DLHC not only is the bond angle of 69° significantly different from the ideal tetrahedral bond angle of 109.47°, but also the ELF shows a complete lack of electron localization in the entire interlayer region.

In order to understand more subtle aspects of the binding in DLHC we investigate the total charge difference $\Delta\rho$, between DLHC and two isolated SLHCs, shown in Figs. 2a and 2b. Evidently, a small amount of charge transfer from Ga to As across the interface has taken place. It results in a binding energy of 1.01 eV/(1x1) calculated at the HSD level. While this value is only a third of the standard Ga-As bonding strength of 3 eV/(1x1), it is 3 times of the interlayer vdW energy (see below). This type of binding is characteristic of dative bond that may be described by a level repulsion between the high-lying empty state of cation and the low-lying doubly occupied state of anion [20], as schematically depicted in Fig. S1a in the Supplementary Information (SI). The repulsion lowers system total energy by lowering the energy of occupied states.

Table I shows the formation energy $\Delta H_f$ with respect to bulk for 28 DLHCs. Figure 2c shows a systematic trend between $\Delta H_f$ and Phillip's ionicity $f_i$ [21], where a least-square fit yields $\Delta H_f = [3.95 f_i (0.9 - f_i)]^2$. We believe the approximate dome shape here reflects the competition between dipole repulsion in Fig. 1b, which increases with $f_i$, and the stability of lone pairs, which also increases with $f_i$. While charge transfer between cation and anion is essential to satisfy the electron counting model (ECM) [22], the resulting dipoles are heads on, due to the central symmetry of the DLHC, and are hence repulsive.

In addition to the ultra-thin limit of a single DLHC, there is a region of stability in which DLHCs can be stacked to form a layered material. Figure 3 shows $\Delta H_f$ as a function of layer thickness $n$ for III-V (GaAs), II-VI (HgSe), and I-VII (AgI), respectively. Unlike the formation of DLHC from the dative bonding between two SLHCs, here the chemically inert DLHCs are held together by vdW forces. Taking GaAs as an example, the binding energy $\Delta E = E(n+1) - [E(n) + E(\text{DLHC})] = 0.35 \, eV/(1 \times 1)$ is quite insensitive to the number of layers, $n$. This corroborates with the fact that $\Delta\rho$ between DLHCs is negligible if plotted at the same contour interval in Fig. 2 (not shown). To further confirm, we performed separate calculations in which the vdW interaction (implemented via a D3 correction) is turned "off". In these cases we found that the inter-DLHC binding, $\Delta E$, becomes essentially zero. The vdW nature of the system can also be determined by considering the large interlayer spacing between the DLHCs, as can be seen in Fig. S2, SI where the atomic structures for 2 DLHC GaAs, HgSe, and AgI are shown.

As a result, the relative stability of the DLHC layered structure and TB form a universal trend. Namely, while the formation energy of the DLHC layered structure is largely independent of the number of layers, $n$, the truncated bulk shows a $1/r$ energy dependence which approaches that of bulk for large $n$. While the bulk ZB or WZ is lower in energy than that of bulk DLHC, the surface dangling bonds of the stable bulk form become energetically costly at small $n$ [23]. Hence, there is a crossover in the stability between the bulk phase and the DLHC phase, with the DLHC being more stable for fewer numbers of layers, which is visible in Fig. 3. A more thorough comparison is given in Table S1 of the SI where the relative formation energy, $\delta(\Delta H_f) = \Delta H_f(n\ DLHC) - \Delta H_f(n\ \text{bilayer}\ TC)$, is tabulated for the 28 semiconductors. It appears that all $n = 2$ DLHCs are stable except for CuCl. As $n$ increases, III-V DLHCs become unstable first, followed by II-VI, and then by I-VII, DLHCs. At $n = 4$, only two III-V DLHCs, i.e., AlAs and AlSb, are stable. At $n = 8$, one II-VI DLHC, i.e., MgTe, is stable. At $n = 10$, however, two I-VII DLHCs, CuI and AgI, remain to be stable. Transition from unstable TB in WZ to stable $n$-DLHC is barrierless. This result is in line with the report of AlP on AlN [5].

Concerning the fabrication of layered structure, we note that other than the use of a vdW cover as in Ref. [6] and/or a vdW substrate, one may also consider laser thinning of a thicker film (transferred on a vdW substrate). We also note that most DLHCs are more stable than silicene ($\Delta H_f = 1.45$ eV/Si$_2$), which has been successfully experimentally fabricated by using a metal substrate. Table I lists six thermodynamically-stable layered binaries, which are $n = 4$ AlAs, AlSb, and MnS, $n = 6$ MgSe, and $n = 10$ CuI and AgI. Their formation energies with respect to silicone are only 26, 21, 17, 15, 3, and 2%, respectively. Finally, we note that formation of 2D structures is energetically favored over 3D clusters. Taking GaAs in Fig. 3a as an example, $\Delta H_f$ is 1.48 eV per GaAs for a stoichiometric 8 GaAs cluster [24], which is even higher than an unstable SLHC.

Shifting focus from the stability of the DLHCs to their electronic properties, we find that most II-VI and I-VII DLHCs have a larger-than-bulk band gap, in line with the expectation from quantum confinement. Surprisingly, however, most III-V DLHCs have a smaller-than-bulk band gap. Detailed PBE results are given in Table S2, SI. From our earlier discussion (Fig. S1, SI), level repulsion is expected to push the valence band states down, while pushing the conduction band state up, thereby further enlarging the band gap. This understanding is clearly at odds with the results obtained for many of the III-V DLHCs. In fact, in the extreme cases of InSb, InAs, GaSb, GaAs, and HgTe, the gap closes to such a degree that band inversion across the Fermi level occurs, as determined by HSD calculation.

To understand this, we compare the band structures of SLHC (Fig. 4a) and DLHC (Fig. 4c) for GaAs. As a simplification, we ignore SOC splitting, which is typically less than 50 meV in SLHC and exactly zero in DLHC. We note that, before the formation of DLHC, the SLHC states in Fig. 4a are in fact doubly degenerate because each level has two identical copies (one from $SLHC_1$ and one from $SLHC_2$, see Fig. S3, SI). The formation of the DLHC lifts the degeneracy. Although a splitting between fully occupied (or fully empty states) has little effect in lowering the system formation energy, it is large enough to completely erase the signature of the first-kind splitting discussed in Fig. S1, SI. If we denote the wavefunctions of the degenerate SLHC states by $\psi_1$ and $\psi_2$, the degenerate eigenenergies by $\varepsilon_{SL}$, and the coupling by $\Delta (> 0)$, the splitting can be modeled by $H = \begin{pmatrix} \varepsilon_{SL} & -\Delta \\ -\Delta & \varepsilon_{SL} \end{pmatrix}$. The solutions are

$$\varphi_+ = \frac{1}{\sqrt{2}}(\psi_1 + \psi_2) \text{ and } \varphi_- = \frac{1}{\sqrt{2}}(\psi_1 - \psi_2), \tag{1}$$

with well-defined wavefunction character (WC) $\chi_2$: when $\chi_2 = (+)$, $\varphi = \varphi_+$ is a bonding state; when $\chi_2 = (-)$, $\varphi = \varphi_-$ is an antibonding state. Since the SLHC states also have their own WC, denoted here as $c_1$, the overall WC of the DLHC states is thus given by a direct product $c_2 = \chi_2 \otimes c_1$. Incidentally, DLHC also has a parity denoted here as $P_2 = +$ for even and $P_2 = -$ for odd. In our choice of atomic origin, $c_2 = P_2$. We find that $c$ is a much better descriptor than $P$, as $c$ does not depend on the crystal symmetry.

Using the WC, we are able to perform a mapping between the $n^{\text{th}}$ band in Fig. 4a and the $n'^{\text{th}}$ band in Fig. 4c, as detailed in Fig. 4b. (Actual mapping involves the projection and identification of individual states, which is summarized in Fig. S3, SI.) The results confirm unambiguously that level splitting is the origin for band inversion in DLHC GaAs. Usually, such an inversion is an indication for TI. However, this is not the case here, as the calculated $Z_2 = 0$. One may notice that, unlike a standard TI where a band inversion takes place between states of opposite parities, here the inversion is a result of level splitting of otherwise non-inverted bands, i.e., bonding state at the valence band maximum (VBM) [$c_1 = (+)$] and antibonding state at the conduction band minimum (CBM) $[c_1 = (-)]$. Since, upon level splitting, the bonding state $[\chi_2 = (+)]$ usually has a lower energy than the antibonding state $[\chi_2 = (-)]$, the newly-formed VBM should have $P_2 = \chi_2 \cdot c_1 = (+)(-) = -$ and the newly-formed CBM should have $P_2 = \chi_2 \cdot c_1 = (-)(+) = -$. The inversion thus happens between two states of the same parity, which cannot affect topological properties so $Z_2 = 0$.

Note that the above discussion applies to DLHC band edge states of any semiconductor regardless if there is a band inversion. Hence, optical transition crossing the band gap at Γ should be dipole forbidden for all of them, as the VBM and CBM have the same parity. Figure S4 shows, as an example, the results for AlAs. Here, the calculation is done by TDHF based on HSD results, which includes also the exciton effect. Despite a direct gap of 2.0 eV, appreciable optical transition only takes place at $\hbar\omega = 3.14$ eV. A similar behavior was found for p-type transparent conducting oxide, delafossite $CuInO_2$ [25].

We note that going from SLHC to DLHC is a layer doubling process, and the resulting WC is $c_2 = \chi_2 \otimes c_1$. Going from DLHC to 2-DLHC is another layer doubling process which would result in $c_4 = \chi_4 \otimes c_2$. Following the above discussion, a band-inverted 2-DLHC should be a TI with non-trivial $Z_2$. Figure 5 shows the HSD results for GaAs, where, due to the highlighted band inversion, $Z_2 = 1$ indeed becomes non-trivial. The same WC argument applies to optical transitions, so band edge transition should also be allowed for 2-DLHCs. We speculate that this thickness doubling rule may apply to other layered materials whether they are 3D TIs or ordinary semiconductors. It appears that SOC is not a deciding factor here for the observed topological properties; a similar conclusion was reached in our recent study of topological carbon [26].

There are other important consequences too, noticeably the possible formation of exciton insulator in DLHC. Despite that the concept of exciton insulator has been proposed half century ago [27], its experimental realization in 3D materials has been elusive. 2D materials can be different: first, the exciton energy will increase by a factor of 4 due solely to a geometric effect [28]; second, owing to a reduction in dielectric screening at lower dimensions [29], further increase in the exciton energy is expected. A recent example is transition metal dichalcogenides (TMDs) for which the exciton energy can be as large as nearly 1 eV in the case of $MoS_2$ [30]. This exceptionally large value correlates with the relatively large exciton energy in bulk $MoS_2$, of about 50 meV, to result in an enhancement factor of about 20. For the moment, it is still a daunting challenge to calculate exciton energy by TDHF to meV accuracy. To get an estimate, here we apply the same enhancement factor to bulk exciton energy of 0.9 meV for HgTe [31]. The result is 18 meV, which is in fact larger than the band gap of 14 meV for DLHC HgTe.

The fact that 2D exciton energy of HgTe can be substantially less than those of TMDs is also intriguing, because it has recently been shown [32] that one may not be able to ionize dopants in spatially-isolated TMDs at any reasonable temperature, for the too-large impurity-bound exciton energy. Exciton energy in binary semiconductor bulk is often only a couple meV or less. Even the 5.7 meV of GaAs and 10.9 meV of CdTe [31] are significantly smaller than that of $MoS_2$. After the scaling up, therefore, the 2D exciton values should still be modest. Thus, for 2D electronics, traditional semiconductors in their layered structure may still be the best choice.

In summary, first-principles calculations point to a new paradigm to discover a potentially new world of 2D layered materials out of traditionally 3D ones. While in the current case of traditional semiconductors, the recipe is to "bury or hide" chemically active cation sites inside the DLHC, the rule could vary in different class of solids and in different structures such as ZB, rocksalt, or perovskite. Even within the same structure class, results can be orientation dependent, e.g., one may build 2D layered structure out of (001)-orientated semiconductor films. The electronic properties of the layered structures can be markedly different from those of 3D bulk, which is not only critically important for novel applications, but also call for new physical understanding beyond traditional solid state theory.

ML, WYX, and TML were supported by the NSF Award No. DMR-1305293, DW was supported by the NSF Award No. EFRI 2-DARE-1542798, and DHC and SBZ were supported by the US DOE Grant No. DESC0002623. The supercomputer time sponsored by NERSC under DOE contract No. DE-AC02-05CH11231 and the CCI at RPI are also acknowledged.

Table I. Formation energy of DLHC with respect to bulk in eV/(AB molecule). In bold are selective examples of low-energy multiple-layer DLHCs before a truncated bulk slab becomes stable.

| compound | | $\Delta H_f(eV)$ | compound | $\Delta H_f(eV)$ | compound | $\Delta H_f(eV)$ | compound | | $\Delta H_f(eV)$ |
|---|---|---|---|---|---|---|---|---|---|
| AlP | | 0.54 | InAs | 0.57 | ZnSe | 0.42 | CuCl | | 0.23 |
| AlAs | 1-DL | 0.46 | InSb | 0.51 | ZnTe | 0.36 | CuBr | | 0.16 |
| | 4-DL | **0.37** | BeS | 0.35 | CdS | 0.34 | CuI | 1-DL | 0.16 |
| AlSb | 1-DL | 0.41 | BeSe | 0.31 | CdSe | 0.41 | | 10-DL | **0.039** |
| | 4-DL | **0.30** | BeTe | 0.32 | CdTe | 0.35 | AgBr | | 0.19 |
| GaP | | 0.72 | MgSe | 1-DL | 0.35 | HgSe | 0.40 | AgI | 1-DL | 0.16 |
| GaAs | | 0.61 | | 6-DL | **0.21** | HgTe | 0.37 | | 10-DL | **0.028** |
| GaSb | | 0.56 | MgTe | 0.54 | MnS | 1-DL | 0.39 | | | |
| InP | | 0.67 | ZnS | 0.49 | | 4-DL | **0.25** | | | |

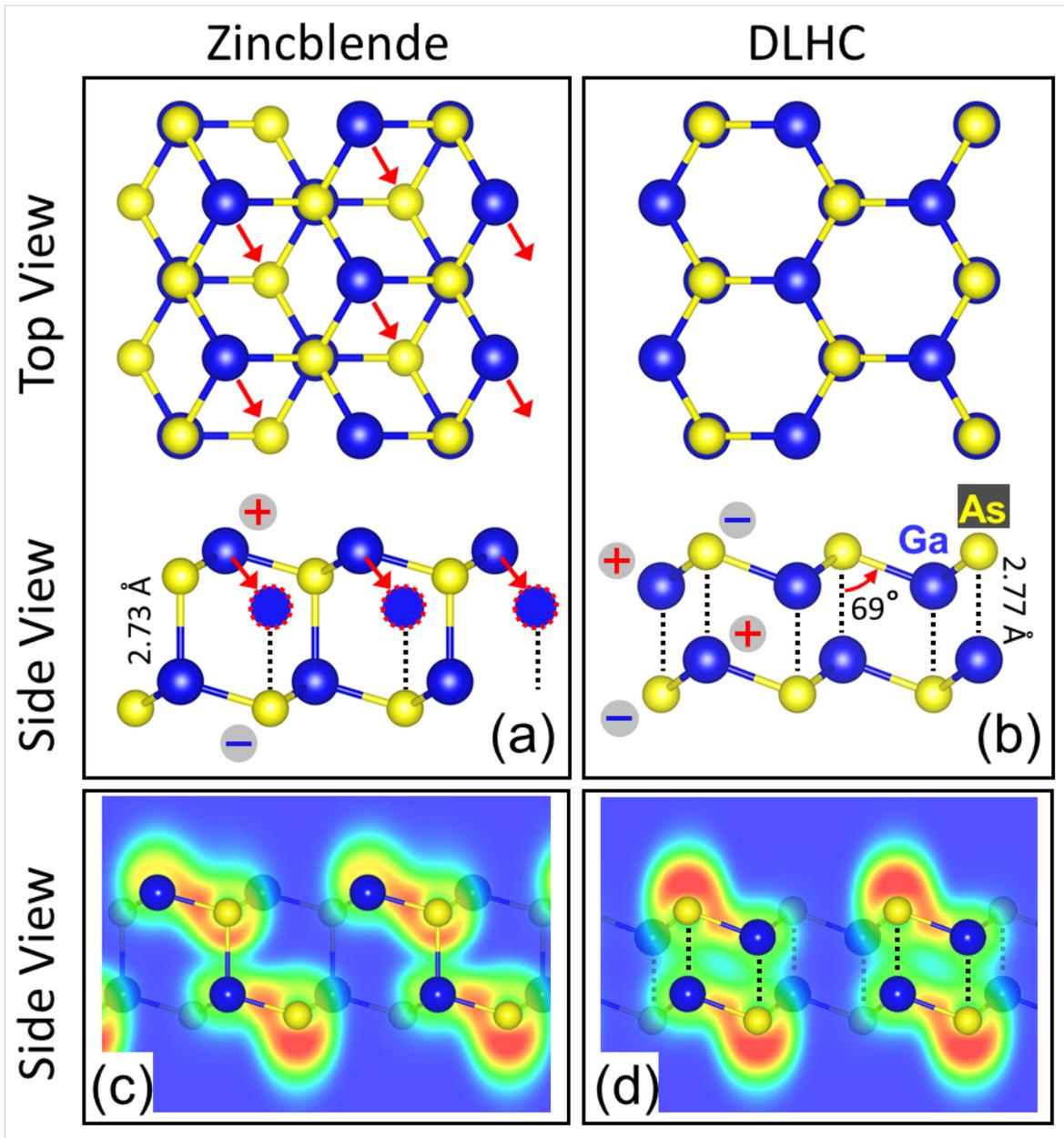

Figure 1. (a) Top and side views of bilayer-thick TB and (b) DLHC GaAs. Red arrows in (a) indicate atomic displacements to form DLHC from TB. Charged plains are denoted schematically by the (+) and (-) signs. (c) and (d) show the corresponding ELFs with contour values ranging from 0 (blue) to 0.8 (red).

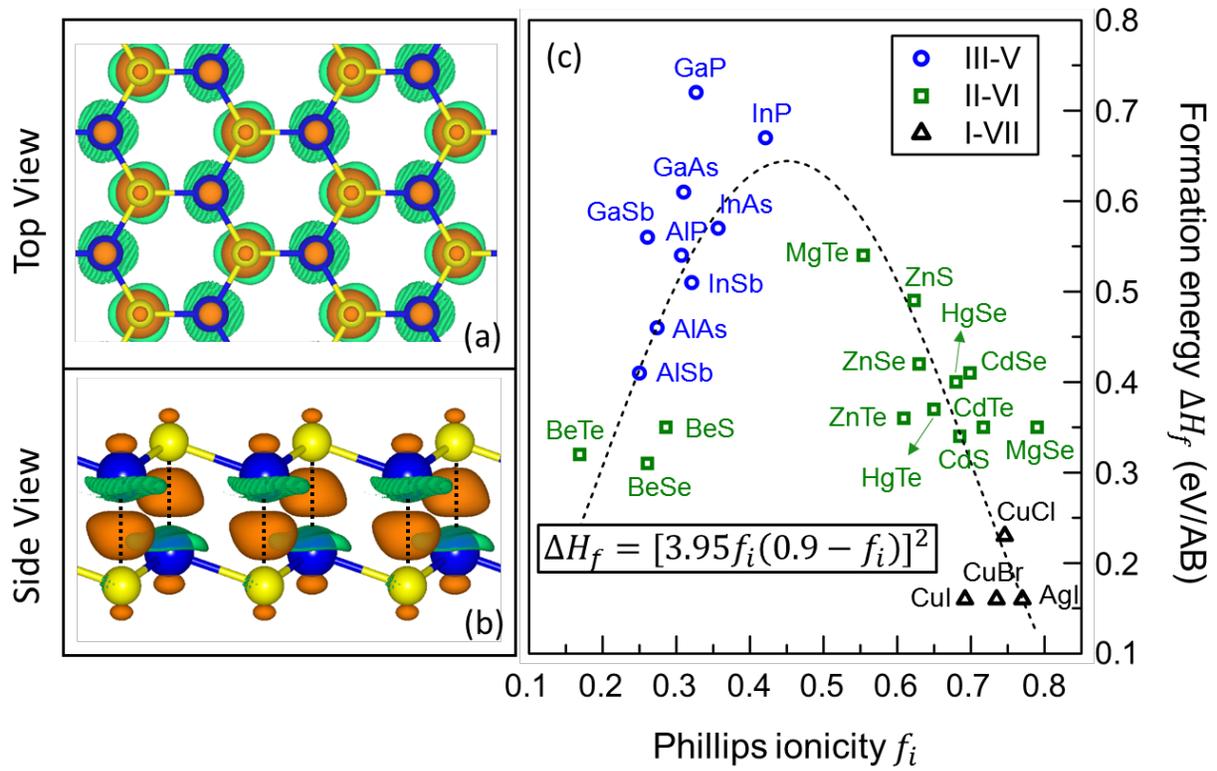

Figure 2. Charge density difference between DLHC and SLHCs for GaAs in (a) top and (b) side view, respectively. Isosurface value is $3 \times 10^{-3} e/\text{Å}^3$; light brown is positive; green is negative. (c) Formation energy as a function of Phillips ionicity.

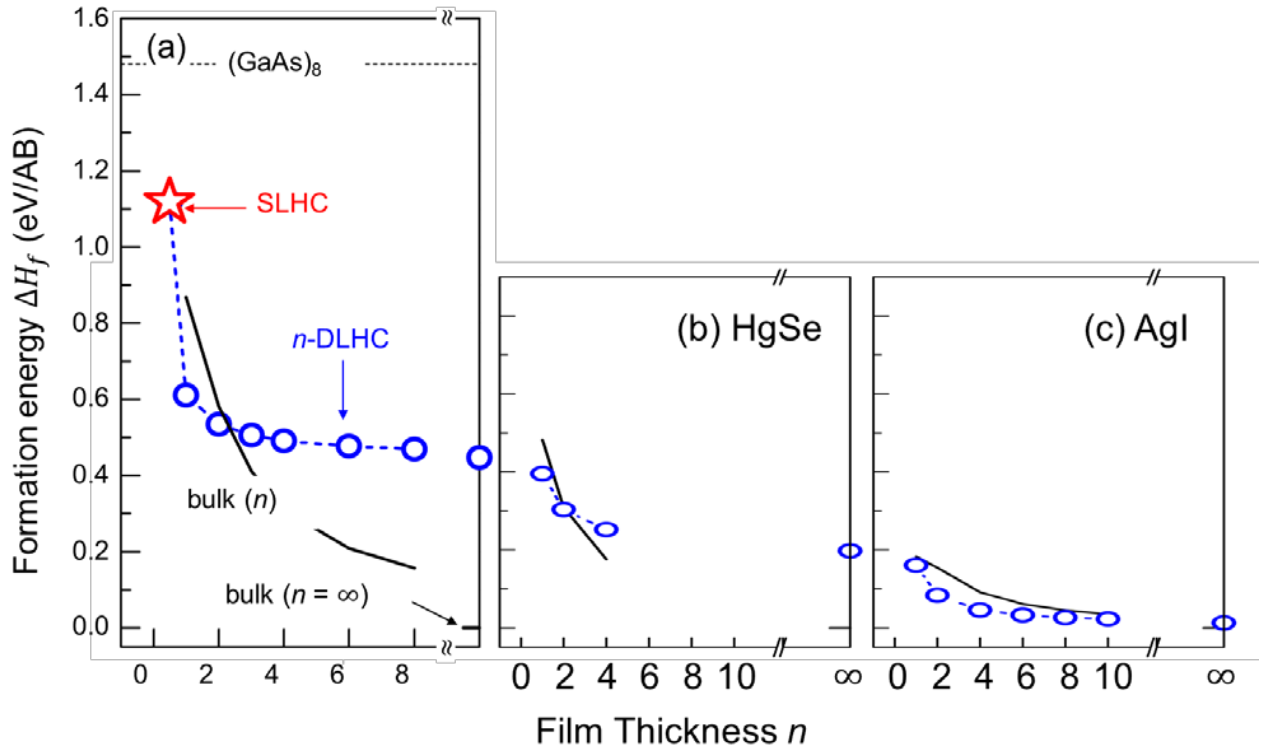

Figure 3. Formation energies of DLHCs (open circles) and TBs (solid lines) as a function of layer thickness for (a) GaAs, (b) HgSe, and (c) AgI. In (a), the formation energy of SLHC and a $(GaAs)_8$ cluster is also shown.

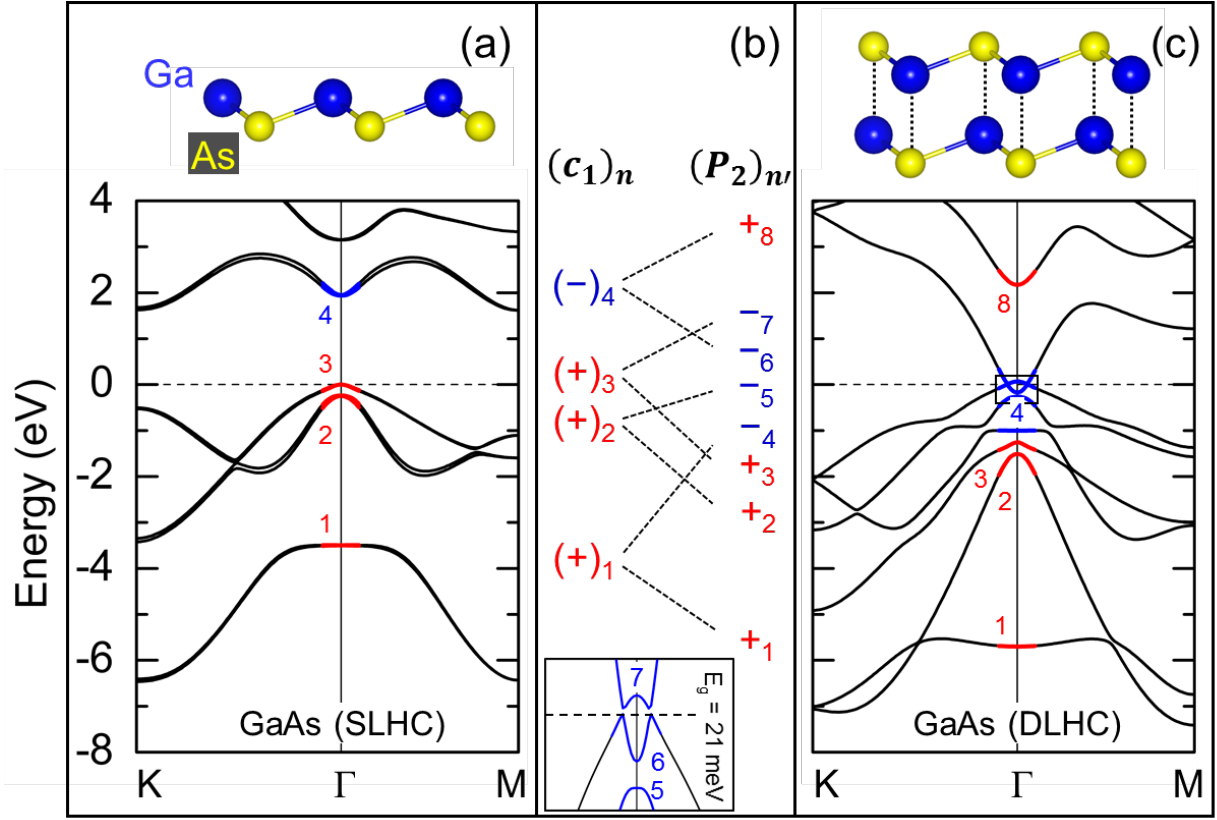

Figure 4. Atomic and band structures of GaAs by HSD. (a) SLHC. (b) Character $(c_s)_n$ of the *n*-th wavefunction of SLHC and parity $P_{n'}$ of the *n'*-th wavefunction of DLHC at Γ (see main text for definition). (c) DLHC. Inset at the bottom of (b) is a blowup of the framed area in (c) showing band inversion. Energy zero is at the valence band maximum.

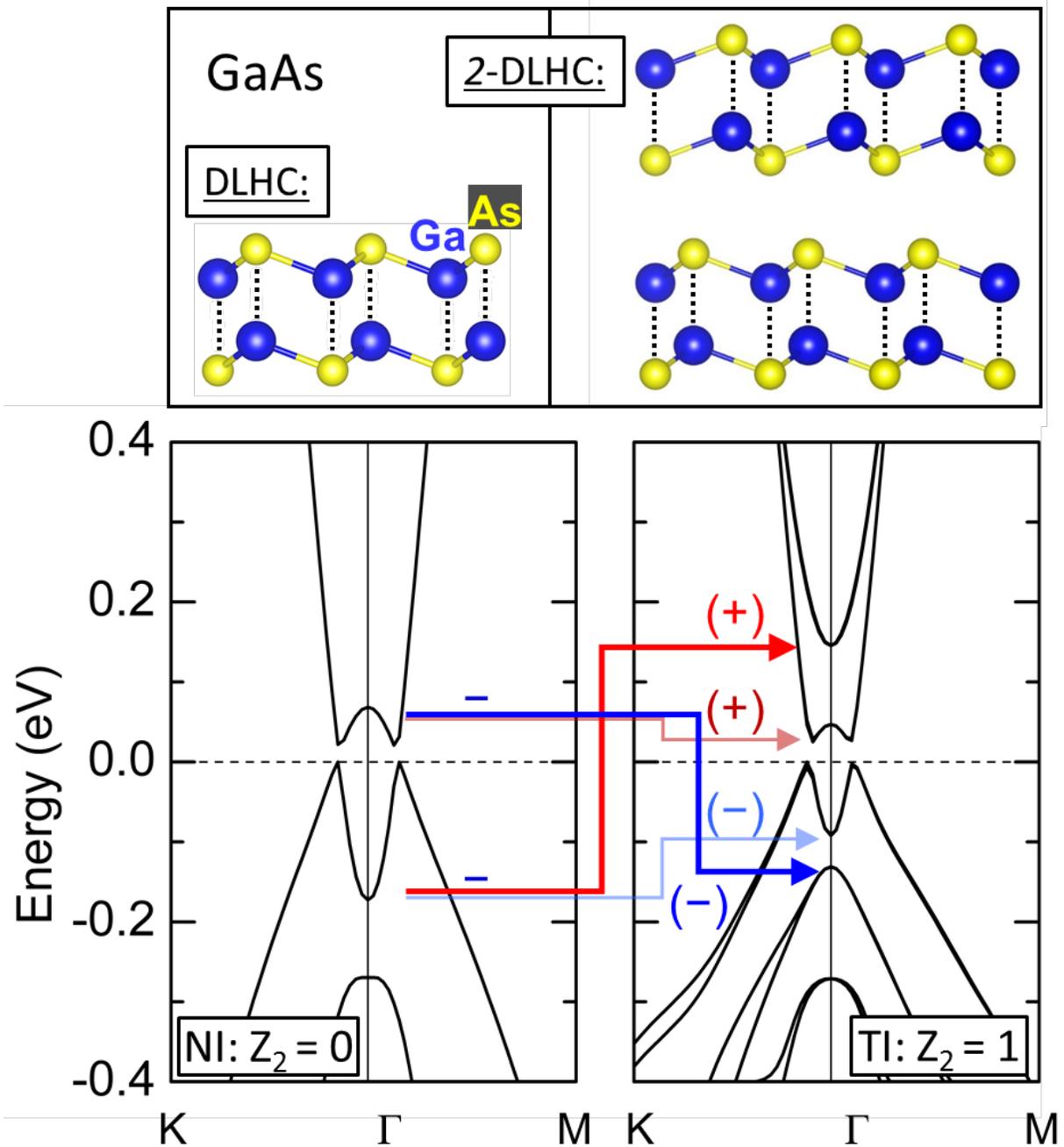

Figure 5. Atomic and band structure of GaAs. Left is DLHC and right is *2*-DLHC. A second band inversion (marked by the crossing between the thickened red and blue arrows) makes the GaAs *2*-DLHC a TI.

**Supplementary Material**

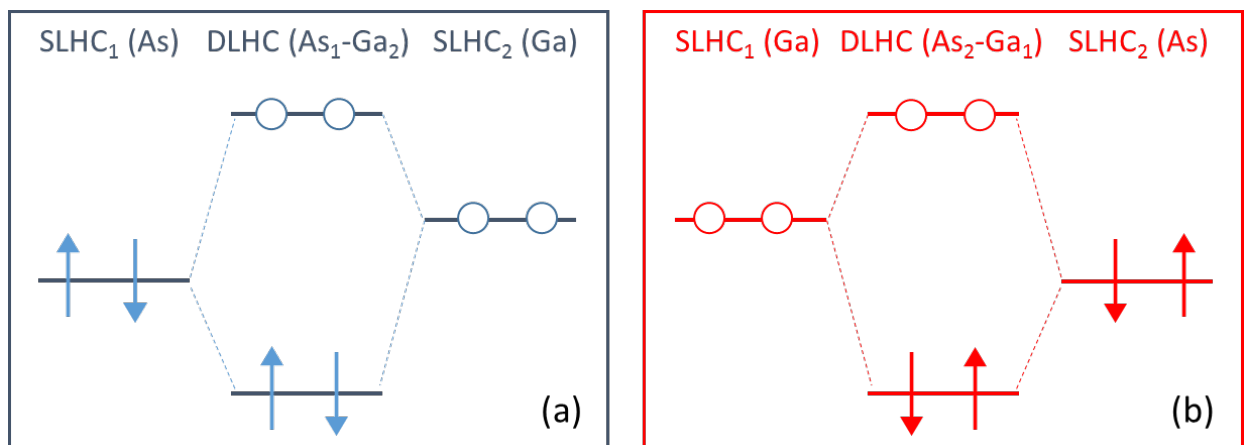

Figure S1. A schematic illustration of level repulsion (a) between occupied As lone-pair state in the first SLHC and empty Ga dangling-bond state in the second SLHC to result in $As_1$-$Ga_2$ dative bonds in DLHC, and (b) between occupied As lone-pair state in the second SLHC and empty Ga dangling-bond state in the first SLHC to result in $As_2$-$Ga_1$ dative bonds in DLHC.

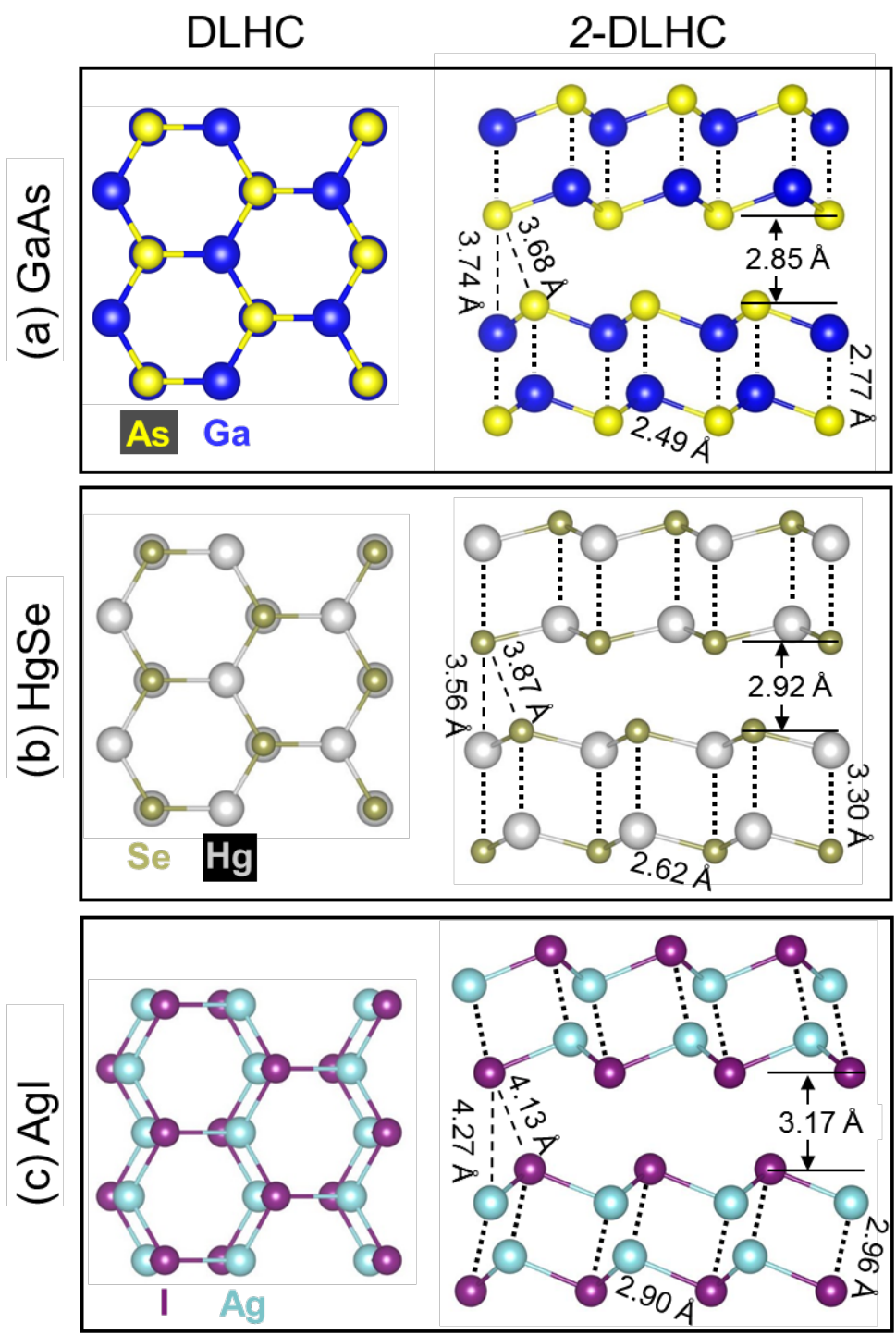

Figure S2. Atomic structures and lattice parameters for DLHC and 2-DLHC (a) GaAs, (b) HgSe, and (c) AgI. Within each panel, a top view of the DLHC is shown to the left. A side view of the 2-DLHC is shown to the right. Noticeable atomic distortion in AgI is present both in DLHC and 2-DLHC.

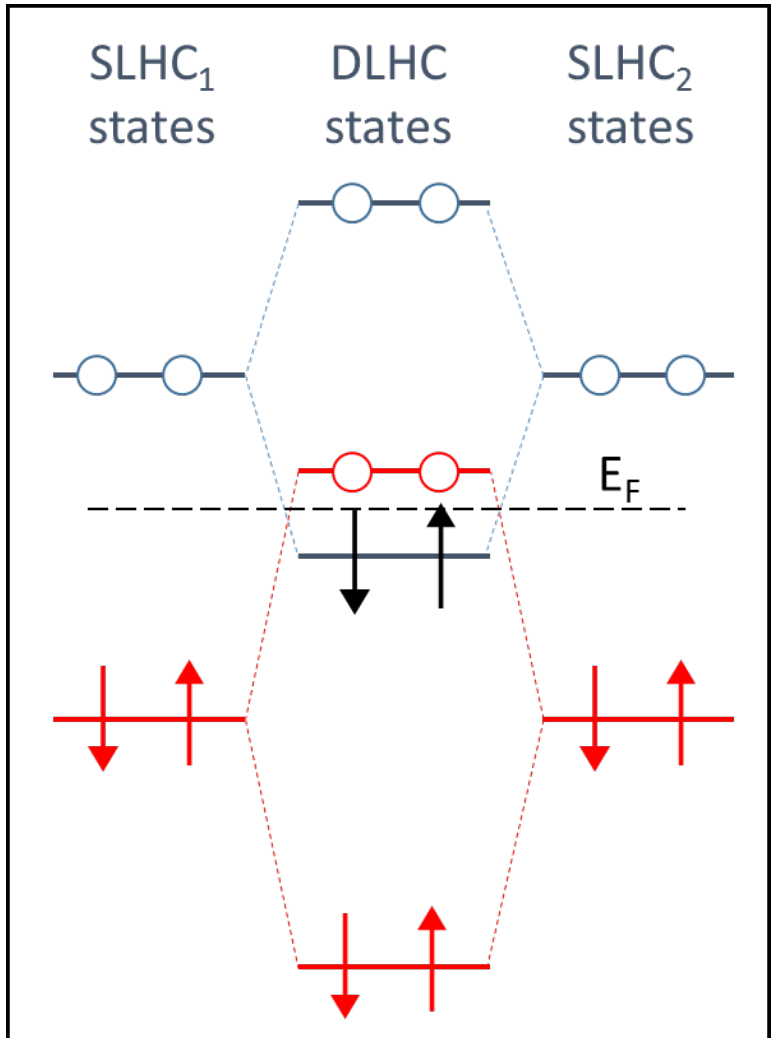

Figure S3. A schematic illustration of level repulsion between degenerate SLHC states. It shows that when the coupling is large enough, band inversion can take place.

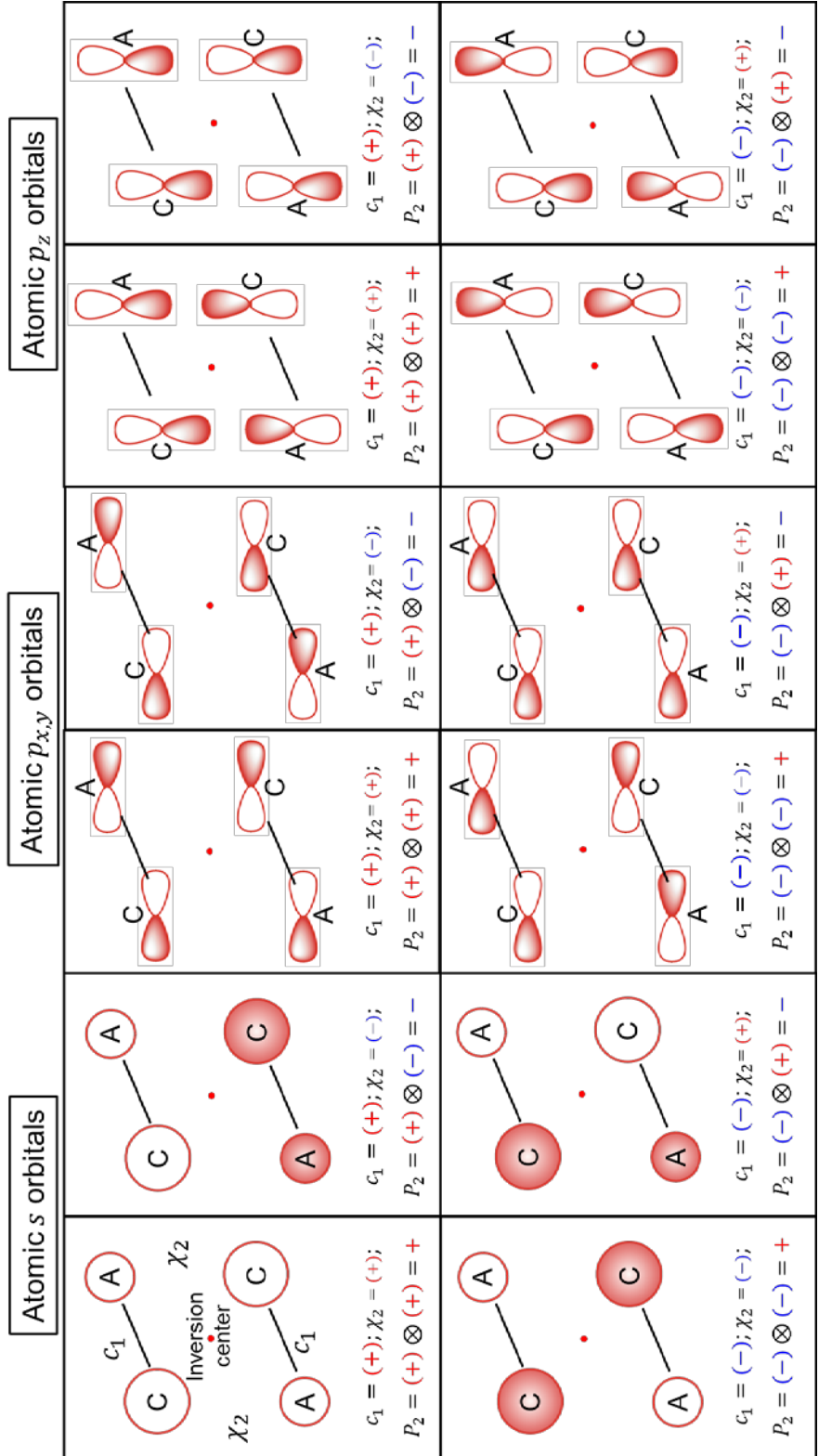

Figure S4. Coupling between SLHC bonding states (top panels) and antibonding states (bottom panels) in a DLHC. C is for cation; A is for anion; open circle is for positive phase; shaded circle is for negative phase. Mapping between SLHC and DLHC near $\Gamma$ in Fig. 4b is shown in green.

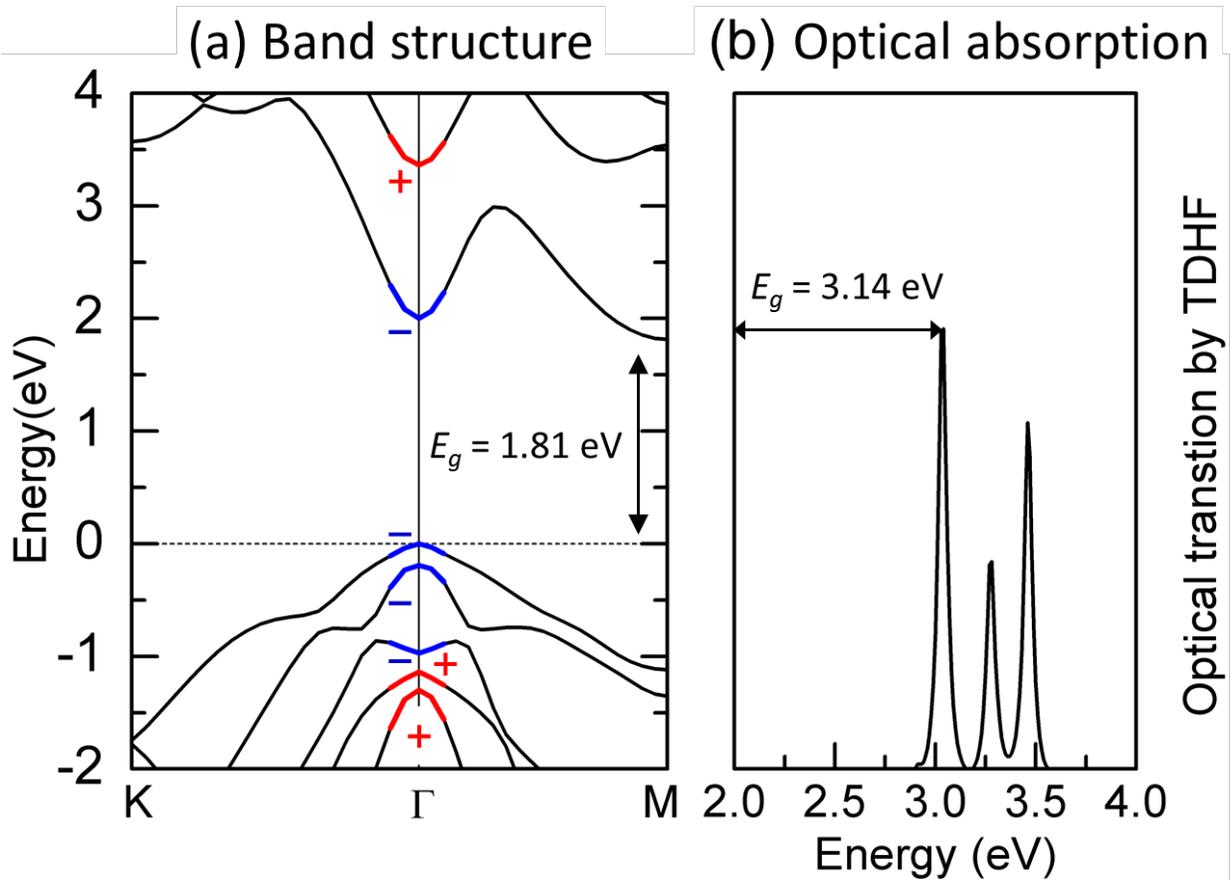

Figure S5. (a) HSD band structure and (b) optical absorption by TDHF for AlAs. Red (+) and blue (−) denote parity of the states at $\Gamma$.

Table S1. Energy difference (meV/AB) between *n*-layer DLHC and truncated bulk of same thickness for 28 semiconductors. Negative means the DLHC has a lower energy. Star by compound name indicates the truncated bulk has a wurtzite structure for which transformation to DLHC is spontaneous.

| Type | Compound | 2-DLHC | 4-DLHC | 6-DLHC | 8-DLHC | 10-DLHC |
|---|---|---|---|---|---|---|
| III-V | AlP | -290 | +45 | | | |
| | AlAs | -300 | -6 | +104 | | |
| | AlSb | -256 | -14 | +79 | | |
| | GaP | -6 | +268 | | | |
| | GaAs | -47 | +181 | | | |
| | GaSb | -34 | +133 | | | |
| | InP | -28 | +209 | | | |
| | InAs | -65 | +135 | | | |
| | InSb | -84 | +85 | | | |
| II-VI | BeS | -243 | -74 | +32 | | |
| | BeSe | -312 | -92 | +2 | | |
| | BeTe | -253 | -55 | +27 | | |
| | MgSe | -324 | -100 | -11 | +35 | |
| | MgTe* | -338 | -127 | -48 | -6 | +19 |
| | MnS* | -292 | -43 | +34 | | |
| | ZnS | -123 | +72 | | | |
| | ZnSe | -148 | +29 | | | |
| | ZnTe | -159 | -1 | +66 | | |
| | CdS* | -104 | +7 | | | |
| | CdSe | -117 | +34 | | | |
| | CdTe | -154 | -10 | +50 | | |
| | HgSe | -6 | +77 | | | |
| | HgTe | -32 | +50 | | | |
| I-VII | CuCl | +23 | | | | |
| | CuBr | -58 | -21 | -5 | +4 | |
| | CuI | -102 | -57 | -31 | -18 | -9 |
| | AgBr | -39 | -17 | -3 | +4 | |

| | AgI | -71 | -46 | -28 | -19 | -12 |

Table S2. Non-spin-orbit band gaps (in eV) of DLHC, bulk, and their differences for 28 AB semiconductors, given by PBE. D and I denote direct gap (at Γ) and indirect gap (at M), respectively.

| Type | Compound | DLHC $E_g$ | bulk $E_g$ | $\Delta E_g$ |
|---|---|---|---|---|
| III-V | AlP | 1.65 (I) | 1.60 (I) | 0.05 |
| | AlAs | 1.36 (I) | 1.47 (I) | -0.11 |
| | AlSb | 0.13 (D) | 1.24 (I) | -1.11 |
| | GaP | 0.35 (D) | 1.64 (I) | -1.29 |
| | GaAs | -0.35 (D) | 0.30 (D) | -0.65 |
| | GaSb | -0.68 (D) | 0.00 (D) | -0.68 |
| | InP | 0.21 (D) | 0.48 (D) | -0.27 |
| | InAs | -.24 (D) | 0.00 (D) | -0.24 |
| | InSb | -0.51 (D) | 0.00 (D) | -0.51 |
| II-VI | BeS | 3.77 (I) | 3.10 (I) | 0.67 |
| | BeSe | 3.03 (I) | 2.62 (I) | 0.41 |
| | BeTe | 1.98 (I) | 1.97 (I) | 0.01 |
| | MgSe | 3.16 (I) | 2.71 (D) | 0.45 |
| | MgTe | 2.95 (D) | 2.61 (D) | 0.34 |
| | MnS | 2.79 (D) | 3.02 (D) | -0.23 |
| | ZnS | 2.66 (D) | 2.16 (D) | 0.50 |
| | ZnSe | 1.85 (D) | 1.30 (D) | 0.55 |
| | ZnTe | 0.96 (D) | 1.23 (D) | -0.27 |
| | CdS | 1.88 (D) | 0.99 (D) | 0.89 |
| | CdSe | 1.53 (D) | 0.57 (D) | 0.96 |
| | CdTe | 1.00 (D) | 0.70 (D) | 0.30 |
| | HgSe | 0.00 (D) | 0.00 (D) | 0.00 |
| | HgTe | -0.16 (D) | 0.00 (D) | -0.16 |
| I-VII | CuCl | 1.34 (D) | 0.60 (D) | 0.74 |
| | CuBr | 1.56 (I) | 0.52 (D) | 1.04 |
| | CuI | 2.07 (D) | 1.24 (D) | 0.83 |
| | AgBr | 1.61 (D) | 1.07 (D) | 0.54 |
| | AgI | 1.94 (D) | 1.34 (D) | 0.60 |